\documentclass[
 reprint,
superscriptaddress,
 amsmath,amssymb,
 aps,
prl,
]{revtex4-2}

\usepackage{graphicx}
\usepackage{dcolumn}
\usepackage{bm}

\usepackage{float}
\usepackage[dvipsnames]{xcolor}
\usepackage{adjustbox}
\usepackage{lineno}
\usepackage{enumitem}
\usepackage{comment}
\begin{document}

\title{Impulsive shock wave propagation in granular packings under gravity}

\author{S. van den Wildenberg}
\email{siet.van\_den\_wildenberg@uca.fr}
\affiliation{Institut Langevin, ESPCI Paris, Université PSL, CNRS, Paris 75005, France}
\affiliation{Université Clermont Auvergne, CNRS, IRD, OPGC, Laboratoire Magmas et Volcans, F-63000 Clermont-Ferrand, France}
\affiliation{Université Clermont Auvergne, CNRS, Laboratoire de Physique de Clermont Auvergne, F-63000 Clermont-Ferrand, France}

\author{X. \ Nguyen}
\affiliation{Institut Langevin, ESPCI Paris, Université PSL, CNRS, Paris 75005, France}

\author{A.\ Tourin}
\affiliation{Institut Langevin, ESPCI Paris, Université PSL, CNRS, Paris 75005, France}

\author{X.\ Jia}
\email{xiaoping.jia@espci.fr}
\affiliation{Institut Langevin, ESPCI Paris, Université PSL, CNRS, Paris 75005, France}


\begin{abstract}
We experimentally investigate the impulsive shock propagation caused by an impact into vertically oriented 3D granular packings under gravity.  We observe a crossover of wave propagation, from sound  excitation at low impact to shock front formation at high impact. One of our findings is a nonlinear acoustic regime prior to the shock regime in which the wave speed decreases with the particle-velocity amplitude due to frictional sliding and rearrangement. Also, we show that the impulsive shock waves at high impact exhibit a characteristic spatial width of approximately 10 particle diameters, regardless of shock amplitude. This finding is similar to that observed in 1D granular chains and appears to be independent of the contact microstructure, whether involving dry or wet glass beads, or sand particles. The final and main finding is that we observe the coexistence of the shock front and the sound waves (ballistic propagation and multiple scattering), separated by a distinct time interval. This delay increases with impact amplitude, due to the increase shock speed on one hand and the decrease of the elastic modulus (and sound speed) in mechanically weakened granular packings by high impact on the other hand. Introducing a small amount of wetting oil into glass bead packings leads to significant viscous dissipation of scattered acoustic waves, while only slightly affecting the shock waves evidenced by a modest increase in shock front width. 
Our study reveals that shock-induced sound waves and scattering play an important role in shock wave attenuation within a mechanically weakened granular packing by impact.
Investigating impact-driven wave propagation through such a medium also offers one way of interrogating a 3D FPUT-like system where nonlinear and linear forces between grains are involved.

.  
\end{abstract}

\maketitle


\section{Introduction}
\label{sect:intro}
Granular media are composed of macroscopic particles that interact through contact forces. They show a wide array of mechanical behaviors that are highly sensitive to external conditions such as confining pressure. Despite their apparent simplicity, these systems exhibit complex collective dynamics, including transitions between fluid-like and solid-like states, and nonlinear responses under stress \cite{Liu_1998,Hecke_2009}. Understanding the transition between these states as well as the mechanisms that govern wave speed, absorption, and scattering is crucial for applications in geophysics and material sciences \cite{Jaeger_1996}.

Wave propagation in granular materials is remarkably rich and complex, due to the intrinsic disorder, nonlinearity, and dissipation present in these systems \cite{Liu_1992,Jia_1999,Goddard_1990}. Unlike homogeneous solids, where elastic waves admit a well-defined continuum description, granular media support a variety of wave modes that strongly depend on their microstructure. Among the key factors governing wave propagation are the inter-particle contact forces and the external confining pressure, both of which significantly influence the stiffness and dynamics of the granular medium. In these materials, the speed of sound is not fixed, but strongly depends on the confining pressure $P_0$. This behavior is rooted in the non-linear nature of the contact forces described by the Hertzian law \cite{Johnson_1985} that relates the normal contact force $f$ between two particles to the deformation $\delta$ at the contact point, $f\sim\delta^{3/2}$, causing the elastic moduli and the sound speed to scale as $P_0^{1/3}$ and $P_0^{1/6}$, respectively. Consequently, in the absence of confining pressure, the medium becomes a 'sonic vacuum' with zero sound speed. In this regime, Nesterenko described the nonlinear behavior of a chain of beads \cite{Nesterenko_1984, Nesterenko_2001}. Assuming Hertzian contact between the grains, he demonstrated that highly nonlinear compression waves can propagate along the chain in the form of solitary or shock waves. 

In 1D granular chains, the existence of these nonlinear waves has been experimentally confirmed \cite{Coste_1997,Lazaridi_1985}. Under long-duration impacts, the system generates a broad wave featuring a well-defined leading front, commonly referred to as a shock wave \cite{Herbold_2007a,Herbold_2007b,Sen_2008}.
Numerical studies on 2D disordered assemblies of frictionless spheres also showed the existence of shock waves, which is in agreement with the theoretical prediction \cite{Gomez_2012a,Gomez_2012b}. These shock waves have two fundamental properties: their amplitude scales with the impact strength, following a power-law relation, while their characteristic spatial extent remains independent of the impact strength \cite{Gomez_2012b}.
Unlike the frictionless particles used in the models, realistic granular materials are 3D and exhibit friction (such as those in experiments), which don't reach an isostatic critical point as pressure approaches zero \cite{Hecke_2009}. In addition to this fundamental difference, experiments involve dissipative grain interactions, which are absent in the simulations.

Despite these differences between model studies and experiments, it has been shown that in disordered granular systems far from the isostatic critical point and with dissipation, shock waves can still be excited and effectively described using the framework developed for conservative shocks \cite{Wildenberg_2013}. More recently, an alternative approach was developed that determines the shock speed based on collision time rather than energy conservation\cite{Clark_2015}. If the initial force law is known, this method is particularly advantageous when restitution losses during particle interactions are important, such as in systems involving soft, deformable particles. 
Moreover, this study experimentally showed the importance of force propagation in heterogeneous contact networks of 2D photoelastic disc packings.
\begin{figure}[ht]
 \begin{center}
\includegraphics[width=0.4\textwidth]{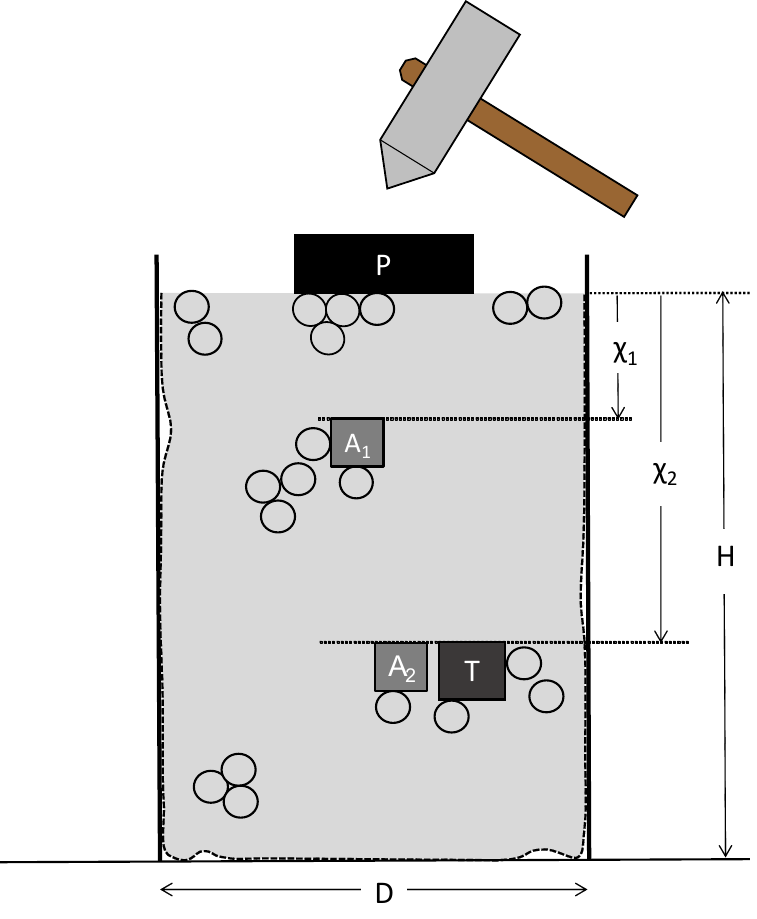}
\caption{Schematic side view of our setup. A container of height~$H=20\, \mathrm{cm}$ and diameter~$D=16\, \mathrm{cm}$ is filled with glass beads. Linear and Shock waves are generated by an impact on a piston~$P$ placed on the top of the beads. Accelerometers~$A_1$ and~$A_2$ are immersed inside the granular medium at distances~$\chi_1=5\, \mathrm{cm}$ and $\chi_2=12\, \mathrm{cm}$ from the source, respectively. A transducer~$T$, also positioned at $\chi_2$, is used to measure the scattered wave.}
\label{fig:fig1}
\end{center}
\end{figure}

However, several important questions remain unanswered. How do friction and dissipation influence the shape of shock fronts? What role does scattering play? Do the scattered waves remain linear?
We address these questions experimentally by studying shock and acoustic wave propagation in granular media subjected to gravitational force. By varying the impact strength, we measure the resulting wave speed and front shape across a broad range of conditions in different granular materials to assess the importance of the microstructure of the particle contacts.
Investigating impact-driven wave propagation through these real granular packings also offers a way to explore the energy exchange between solitary-like shock waves and acoustic oscillations, i.e. the FPUT-like problem (Fermi-Pasta-Ulam-Tsingou), in a 3D, disordered and dissipative system where nonlinear and linear forces between grains are involved \cite{Sen_2008}.

\section{Experimental setup}
Our setup consists of a cylindrical cell with a height of H = 20 cm and a diameter of D = 16 cm, made of paper and with a plastic cover inside. The cell is filled with glass beads (diameter $d=$~3 or 5$\, \mathrm{mm}$), sand particles ($d=3\, \mathrm{mm}$) or glass beads in the presence of a small amount of oil \cite{Brunet_2008}.
To create waves, we use a hammer (Br\"uel \& Kjaer) to impact a teflon piston (diameter~6 $\, \mathrm{cm}$ , thickness~2 $\, \mathrm{cm}$, mass~110 $\, \mathrm{g}$) that is placed on top of the grains. We vary the impact strength over three decades by lightly tapping the piston, resulting in the weakest impacts we can detect, to hitting the piston hard, which generates the strongest impacts. For the strongest impacts, the piston penetrates the packing significantly ($\sim$~5 $\, \mathrm{mm}$) and the particles are ejected from the cell, limiting our impact range. 
We detect wave propagation throughout the granular material by making use of two accelerometers (Br\"uel \& Kjaer) buried in the granular medium at two different distances from the piston ($\chi_1=5\, \mathrm{cm}$ and $\chi_2=12\, \mathrm{cm}$) (Fig.~\ref{fig:fig1}), as well as a broadband piezoelectric transducer also placed at $\chi_2=12\, \mathrm{cm}$. The accelerometers allow us, by integration, to measure the local particle velocities and, by another integration, the local displacements. The piezoelectric transducer is sensitive to higher frequencies, allowing us to measure the scattered wave.
For each experimental condition, we performed two experiments, each consisting of~150 to~200 impacts, which were then merged for analysis. For each experiment, the container was freshly filled and after each experiment the correct position of the sensors was verified. 

\begin{figure*}[ht]
 \begin{center}
\includegraphics[width=1\textwidth]{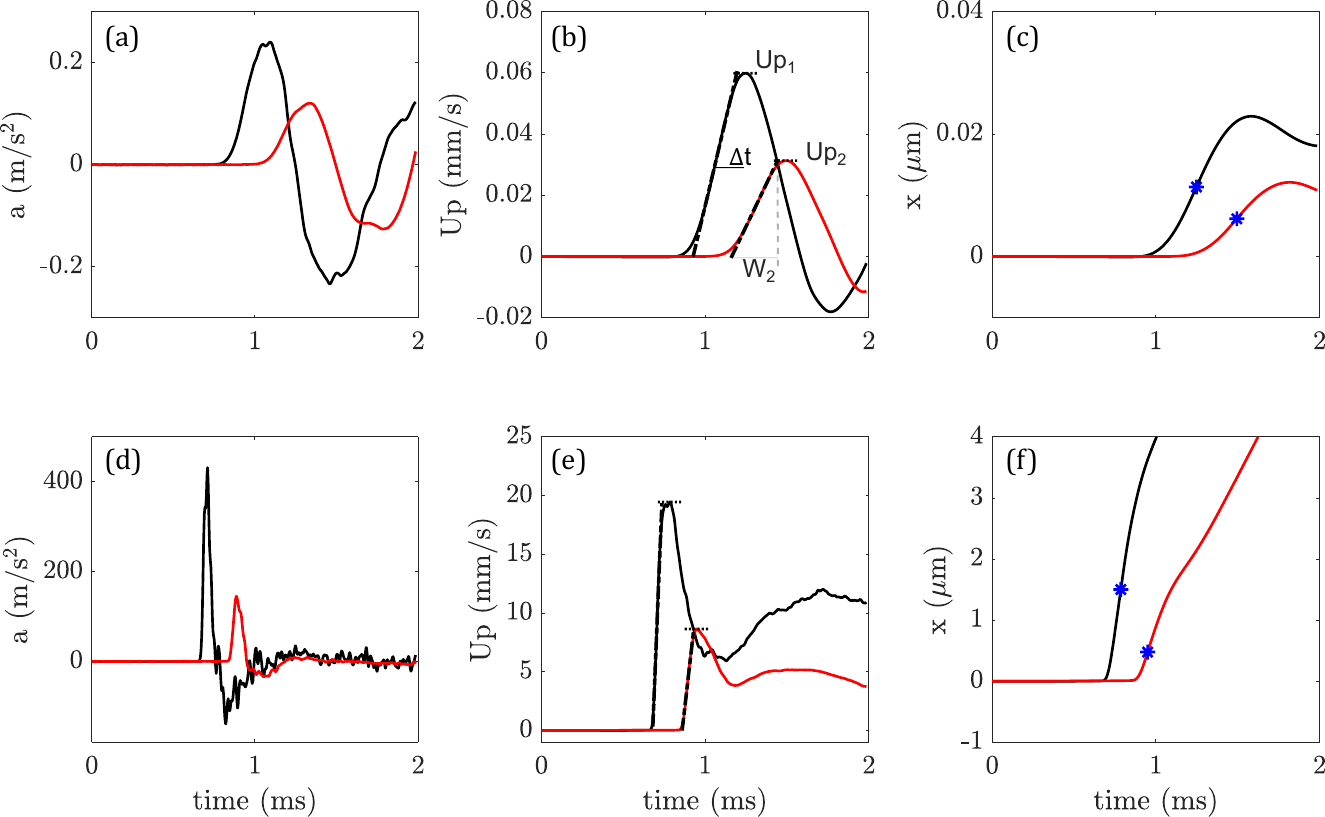}
\caption{Typical signals for low (a-c) and high (d-f) impact amplitude, at $\chi_1= 5\, \mathrm{cm}$ (black) and $\chi_2=12\, \mathrm{cm}$ (red). (a,d) Local acceleration a(t). (b,e) Local particle velocity $U_p(t)=\int_{0}^{t} a(\tau) d\tau$. The peak velocities $U_{p_1}$ and $U_{p_2}$ are indicated, as well as the width (only for $W_2$) and the delay $\Delta t$ between the mid points of the slopes. (c,f) Local displacement $x(t)=\int_{0}^{t} U_p(\tau) d\tau$, stars (blue) indicate the peak times of $U_p$.}
\label{fig:fig2}
\end{center}
\end{figure*}

\section{Shock Propagation}
\subsection{Phenomenology}
Our sensors allow us to extract the time evolution of the local acceleration, velocity, and displacement at locations $\chi_1$ and $\chi_2$ from the source. Representative examples of these profiles, both for a weak and a strong impact measured in glass bead assemblies, are shown in Figure~\ref{fig:fig2}. They qualitatively illustrate important aspects of the phenomenology of the waves we observe. The local accelerations achieved in our impact experiments range from several hundred $\, \mathrm{cm/s^2}$ for very soft strokes up to several hundred $\, \mathrm{m/s^2}$ for the hardest hits. From integrating the accelerations we obtain the local particle velocity; the peak velocities $Up_1$ and $Up_2$ are then defined as the first local maxima in the velocity signals. Subsequently, integrating the particle velocity signal yields the particle displacements (Fig.\ref{fig:fig2}) where the measured maximal value is on the order of several micrometers. 

We may estimate the particle displacement by invoking that, in the shock regime, the peak kinetic and potential energies are on equal terms \cite{Gomez_2012a,Wildenberg_2013}. The peak kinetic energy is $(1/2)m~U_p^2$ with grains of $R=d/2\sim~1.5\, \mathrm{mm}$ and $m\sim 3.5~10^{-5}\, \mathrm{kg}$. The peak potential energy for particles interacting via Hertzian contact forces is $(2/5)K~\delta^{5/2}$, where $K=2E^*R^{1/2}\sim 4~10^{6}\, \mathrm{N/m}$, using a typical value of $E^*= 50 \, \mathrm{GPa}$. 
By assuming that the peak potential energy of $z$ contacts equals the peak kinetic energy and taking $Up\sim 3 \, \mathrm{mm/ s}$ we arrive at a total displacement of $\delta=(5~mUp^2/(4zK))^{2/5}\sim 0.9 \, \mathrm{\mu m}$, with $z=5$ (a typical value for dense packing of frictional spheres). Note that $\delta$ includes both the deformation due to static pressure $\delta_0$ and the dynamic part $\delta_D$ due to the shock. However, we neglect the contribution from static pressure, as it is very small: $\delta_0\sim(3\pi R^2P_h/(4E^*R^{1/2})\sim 0.016\, \mathrm{\mu m}$, where the hydrostatic pressure $P_h=\rho_mg~\chi_1 \sim 750\, \mathrm{Pa}$, with $\rho_m=1500 \, \mathrm{kg.m^{-3}}$.
Assuming the shock front consists of around 10 particles (as discussed in Section Shock Width), the cumulative displacement associated with the shock front is expected to be around $9 \, \mathrm{\mu m}$, which is reasonably comparable to the observed displacement ($\sim 2 \, \mathrm{\mu m}$) at the leading edge of the shock. This implies that the macroscopic rearrangements of the particle positions remain negligible in our experiments.

\begin{figure}[ht]
 \begin{center}
\includegraphics[width=0.5\textwidth]{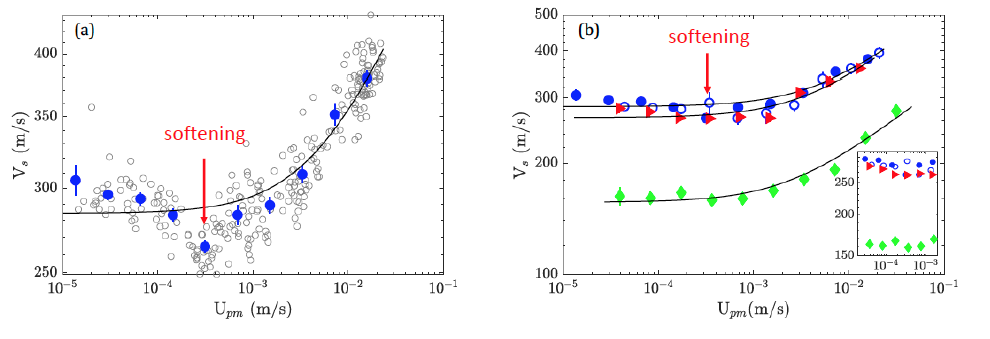}
\caption{Front propagation speed as a function of the front particle-velocity $U_{pm}$. (a) $V_s$ versus $U_{pm}$ for all the pulses measured in assemblies of glass beads with diameter $d=~3\, \mathrm{mm}$ (open gray symbols). The mean value is calculated over multiple pulses (solid dots), and the error bar represents the standard deviation. (b) $V_s$ versus $U_{pm}$ for  $d=~5\, \mathrm{mm}$ glass beads (open blue symbols), $d=~3\, \mathrm{mm}$ glass beads with oil (red right-pointed triangles), sand (green diamond). The data from (a) are also included (solid dots). Packings of glass beads show a softening ($\sim10 \%$) of the sound speed at $U_{pm}< 5~ 10^{-4} \, \mathrm{m/s}$ (inset). Error bars represent the standard deviation. Fits (solid lines) correspond to functions of the form $V_s=C (1+ Up_m/Up_f)^\frac{1} {5}$. 
}
\label{fig:fig3}
\end{center}
\end{figure}

\subsection{Shock and sound propagation speeds} 
To investigate the nature of the waves excited in our system, we determined the propagation speed~$V_s$, which we define as the speed of the wave front. The propagation speed is determined from the delay in arrival time, $\Delta t$, determined at the 50\% amplitude point of $Up$ measured at positions $\chi_1$ and $\chi_2$. To accurately determine when the peak magnitude of 50\% is reached, we apply linear fits over the interval from 20\% to 80\% of the leading slope of the particle velocity signals. $V_s$ is plotted as a function of $Up$ in a so-called Hugoniot plot (Fig.~\ref{fig:fig3}). Due to attenuation, $Up_2 < Up_1$. To characterize the impact strength consistently, we use the geometric mean, $Up_m=\sqrt{Up_1Up_2}$ \cite{Wildenberg_2013}. We observe that at low impact strength, $V_s$ does not depend on the local particle velocity with a value expected to be close the sound speed, while at high impact strength, $V_s$ increases with the velocity of the particles, a behavior characteristic of granular shocks.

More specifically, the data for glass beads, sand particles, and glass particles in oil are well described by the relation $V_s=C (1+ Up_m/Up_f)^\frac{1} {5}$,  which captures both the plateau at low impacts and the power-law increase at higher impacts \cite{Gomez_2012a,Wildenberg_2013}.
In assemblies of glass beads, the transition from linear to shock waves occurs at $Up_f\sim 4.8\, \mathrm{mm/ s}$. This crossover remains largely unchanged in packings where oil is present at the contacts, as well as in sand particle packings ($Up_f\sim 3.6\, \mathrm{mm/ s}$ and $Up_f\sim 2.4\, \mathrm{mm/ s}$, respectively). The similarity in $Up_f$ values suggest that the transition to the shock regime is not strongly influenced by the microstructure of the contact, but is mainly governed by the elastic Hertzian interaction between the particles, as predicted by theoretical models \cite{Gomez_2012a,Clark_2015}. Note that the speed of sound is lower in sand packings compared to glass bead packings, due to macroscopic friction arising from the angularity of the particles. This friction enables the formation of a stable packing with relatively low coordination number and thus a lower speed of sound \cite{Wildenberg_2015}.

\begin{figure}[ht]
 \begin{center}
\includegraphics[width=\linewidth]{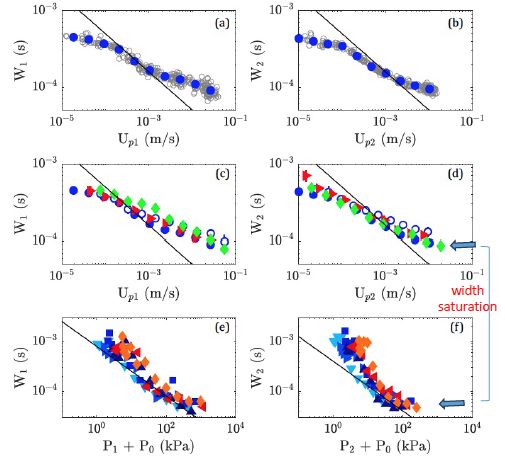}
\caption{Impulsive front width as a function of pulse amplitude at positions $\chi_1$ and $\chi_2$. 
(a) $W_1$ versus $U_{p1}$ for all pulses measured in packings of $d=3\, \mathrm{mm}$ glass beads. Open gray symbols represent individual impulses and solid dots indicate the mean over multiple impulses. (b) same as (a), but for $W_2$ versus $U_{p2}$. (c) $W_1$ versus $U_{p1}$ for $d=~5\, \mathrm{mm}$ glass beads (open blue symbols), $d=~3\, \mathrm{mm}$ oil-wet glass beads (red right-pointed triangles), sand (green diamond), the data from (a) is included (solid dots). (d) Same as (c), but for $W_2$ versus $U_{p2}$. (e) $W_1$ versus the total pressure $P_1+P_0$ using the experimental data from ref.\cite{Wildenberg_2013}, with different colors representing different confining pressures $P_0$. (f) Same as (e), but for $W_2$ versus $P_2+P_0$ from ref.\cite{Wildenberg_2013}. Error bars represent standard deviations and solid lines indicate the relation $W\sim~1/\sqrt{A}$ predicted by equations (1) and (2) with $A$ the impulse amplitude in the weakly nonlinear regime.}
\label{fig:fig4}
\end{center}
\end{figure}

Importantly, at intermediate impact strengths prior to the onset of the shock regime, we observe an approximately $10\%$ reduction in sound speed (Fig.\ref{fig:fig3}b,inset), indicating a softening of the granular packing in the weakly nonlinear regime due to the frictional sliding between grains and associated rearrangements of the contact-force networks without significant grain motions\cite{Jia_2011}, as mentioned above. In contrast, no such softening is observed in sand packings, likely due to the interlocking of the angular grains.

\section{Shock Width}
The small elastic deformations observed at the leading edge of the shock wave, along with the reported weak log type dissipation \cite{Wildenberg_2013}, suggest that attenuation of impulsive shock fronts is primarily driven by the geometric spreading and inelastic collisions rather than scattering. However, as shown in Fig.\ref{fig:fig2}, the shock front steepens with increasing impact strength, and the resulting higher frequencies would likely lead to more scattering. To explore this further, we analyze the impulse duration, characterized by $W$, which we derive from the slope (temporal/spatial) of $Up$ (Fig.~\ref{fig:fig2}). In Fig.~\ref{fig:fig4}, we observe the decrease of $W$ as a function of $Up$. We have also included $W$ obtained from data found in ref.\cite{Wildenberg_2013} (Fig.~\ref{fig:fig4}e,f) which were found to decrease as a function of the total pressure (calculated as the sum of the dynamic and confining pressures). We find that impulsive front duration is similar for sand and glass particles of comparable size with only a slight influence of wetting at the contact (see discussions below). 
The impulse width also appears independent of confining pressure, indicating that it is not affected by the contact microstructure.

\begin{figure*}[ht]
 \begin{center}
\includegraphics[width=0.8\textwidth]{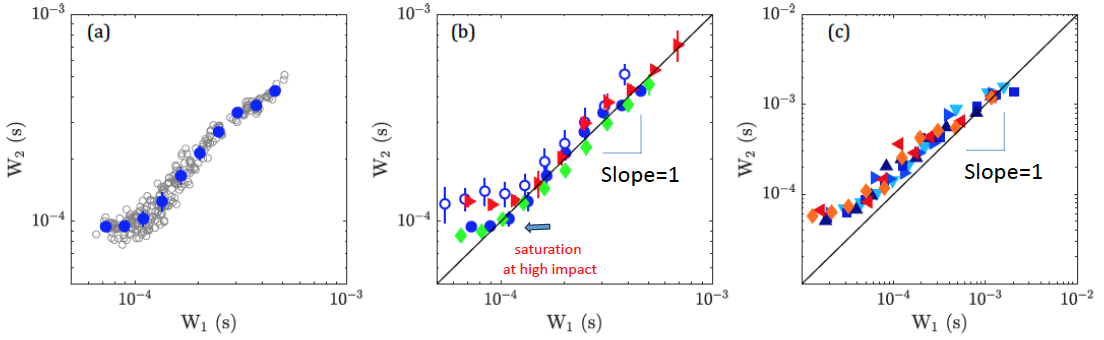}
\caption{Scatter plots of the pulse front width $W_2$ at $\chi_2$ and $W_1$ at $\chi_1$. (a) $W_2$ versus $W_1$ for all pulses in packings of $d=~3\, \mathrm{mm}$ glass beads (open gray symbols). Mean values are shown as solid dots. (b) Same as (a), with additional data for  $d=5\, \mathrm{mm}$ glass beads (open blue symbols), $d=3\, \mathrm{mm}$ oil-wet glass beads (red right-pointed triangles), and sand particles (green diamond). (c) $W_2$ versus $W_1$ from experiments reported in ref.\cite{Wildenberg_2013}. Error bars represent standard deviations and solid lines indicate the proportional relationship  $W_2\propto W_1$.}
\label{fig:fig5}
\end{center}
\end{figure*}

To further explore the width of wave fronts generated by impact, we cross-plot $W_2$ versus $W_1$ in (Fig.\ref{fig:fig5}). Despite of the important amplitude attenuation of $U_{p2}$ versus $U_{p1}$ (Fig.\ref{fig:fig2}), we find that the data of $W_2$ versus $W_1$ collapsed onto a line with a slope close to 1, indicating little change in the width. However, at the highest impacts with short impulse width $W_1<0.1\,\mathrm{ms}$, $W_2$ no longer decreases linearly with $W_1$, but instead saturates to a characteristic time such that $W_2 < W_1$ (Fig.\ref{fig:fig5}b). This saturation effect is particularly clear for the impulse widths measured in the experiments in ref.\cite{Wildenberg_2013} on weakly compressed granular packings at higher impact levels. These experiments show a more clear deviation from linear as $W_1\sim 0.01\,\mathrm{ms}$. As will be discussed in the next section, this saturation of impulse width is likely not related to wave scattering.


Overall, the data in Fig.\ref{fig:fig5} reveal three distinct regimes: (i) a linear regime at low impact strengths, (ii) a weakly nonlinear regime at intermediate impacts, and (iii) a shock regime at high impacts. In the linear regime, the width can be estimated from  $(V_s\cdot W)/d \ (\sim~300\, \mathrm{m/s}\cdot5~10^{-4}\, \mathrm{s}/3~10^{-3}\, \mathrm{m} )\sim$~15$\, \mathrm{cm}$, corresponding to 50 particles. In this regime, the width is constant and does not depend on the impact strength~$Up_m$. 
In the weakly nonlinear regime, the impulse width decreases with increasing impact strength. This decrease is described by the Korteweg-de Vries (KdV) equation:
 \begin{equation}
{\frac{dU_p}{dt}+ U_p\frac{dU_p}{d\delta}+ \frac{d^3U_p}{d\delta^3}=0}
\label{eq:KdV}
\end{equation}
The solution of this equation is well-known and has the form: 
 \begin{equation}
 {U_p(\delta,t)= A ~\mathrm{sech}^2 \left(\frac{\delta-ct}{\Delta} \right)}
\end{equation} 
where $A$ represents the peak $U_p$ and the width scaling  $\Delta\sim 1/\sqrt{A}$ supports the observed decrease in impulse width as the amplitude increases (Fig.\ref{fig:fig4}). Interestingly, in the shock regime, we observe that the impulse width saturates and becomes independent of the shock wave amplitude. We estimate that the width of the pulses in this regime is $(V_s\cdot W)/d \ (\sim400\, \mathrm{m/s}\cdot 10^{-4}\, \mathrm{s}/3~10^{-3}\, \mathrm{m})\sim$~10 particles, much larger than the particle size. 
Consistently, in packings of 5$~\mathrm{mm}$ glass beads the temporal width $W$ is slightly smaller and appears to saturate at somewhat lower values of $Up$. This leads again to the characteristic spatial size of the shock of $W\cdot V_s/d \ (\sim400\cdot 1.2~10^{-4}/5~10^{-3})\sim 10$ particles. 
The characteristic spatial width of the shock wave, approximately 10 particle diameters and independent of shock amplitude, agrees with findings from numerical 2D models of frictionless particles \cite{Gomez_2012b}. This relatively large width (wavelength $\lambda_{\text{shock}} \sim 10d$) also rules out significant contributions from scattering (see below), indicating that frictional dissipation and scattering have minimal influence on shock width.



\begin{figure}[ht]
 \begin{center}
\includegraphics[width=0.5\textwidth]{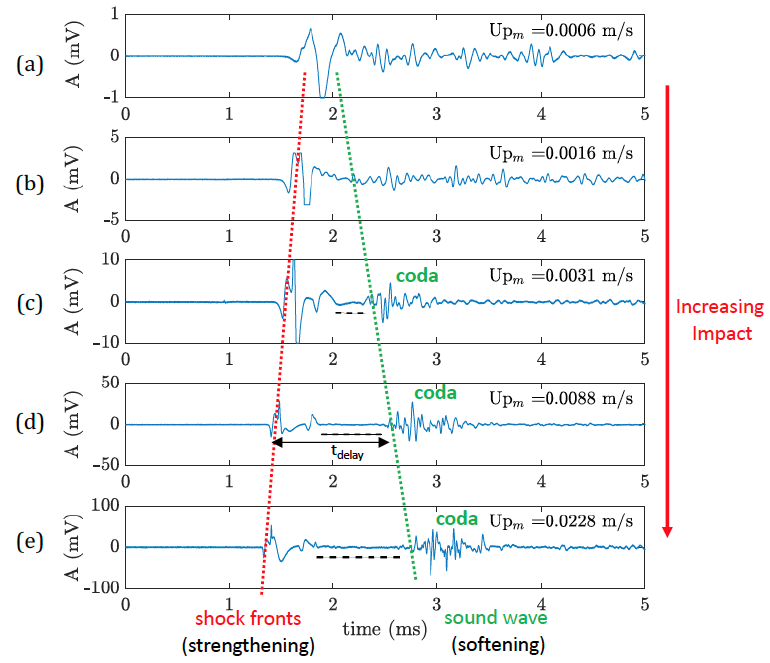}
\caption{Typical signals recorded by the piezoelectric transducer at position $\chi_2$ in dry packings of $d=~3\, \mathrm{mm}$ glass beads for different impact strength. (a) $U_{pm}=0.6 \, \mathrm{mm/s}$. (b) $U_{pm}=1.5 \, \mathrm{mm/s}$. (c) $U_{pm}=3.1 \, \mathrm{mm/s}$. (d) $U_{pm}=8.8 \, \mathrm{mm/s}$. (e) $U_{pm}=22.8 \, \mathrm{mm/s}$.  The dashed lines (black) indicate the onset of a silent period (time delay) between the shock fronts (red dotted line) and sound arrivals consisting of ballistic (green dotted line) and scattered coda waves for higher impacts.}
\label{fig:fig6}
\end{center}
\end{figure}

\section{Coexistence of shock and scattered sound waves}
To investigate the attenuation and energy transfer of shock waves, 
we study our experimental data collected by the piezoelectric transducer particularly sensitive to the elastic deformation of granular packings. Fig.\ref{fig:fig6} shows typical waveforms detected by the transducer in glass bead packings at $\chi_2$ for different impact strength. A striking observation is that, at higher impacts when shock fronts are formed, a distinct "silent" period or delay ($t_{delay}$) develops between the shock waves and the sound waves arrivals, consisting of ballistic and scattered coda waves (see below the sound speed determination). This delay increases with impact amplitude, likely due to the strengthening of shock speed and the softening of the sound speed in mechanically weakened granular packings by high impact. Interestingly, such a silent period has also been observed in simulations of a 1D bead chain with limited mass dispersion \cite{Manjunath_2012}, and arises from the faster speed of the leading pulse compared to the scattered wave. As impact strength decreases, the leading pulse slows down, and the silent period vanishes.

\begin{figure*}[t]
 \begin{center}
\includegraphics[width=0.8\textwidth]{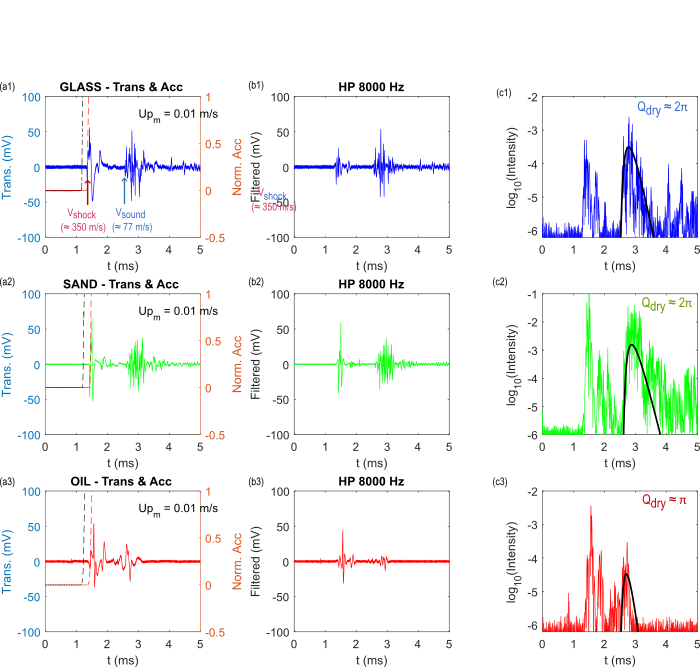}
\caption{(a) Shock and sound waves generated by high impact $U_p\sim \mathrm{10 mm/s}$ and detected by a piezoelectric in dry packings of glass beads and sand particles as well as in a wet packing of glass beads (all grain sizes  $d\sim \mathrm{3 mm}$). (b) Filtered waveforms by a high-frequency pass filter $f_c\sim \mathrm{8 kHz}$. (c) Smoothed intensity profiles over one realization of the corresponding wave signals in (b). Solid line curves (black) correspond to the diffusion solutions to Eq (3).}
\label{fig:fig7}
\end{center}
\end{figure*}

We further observe the coexistence of shock and sound waves at high impact in various granular packings, dry, oil-wet or consisting of nonspherical sand particles (Fig.\ref{fig:fig7}a). 
With the differential time $\Delta {t}$ measured respectively by accelerometers at $\chi_1$ and $\chi_2$ (Fig.\ref{fig:fig2}e), we determine accurately as above a shock speed $V_{shock}\sim 350 \, \mathrm{m/s}$ for the particle velocity $U_{pm}\sim10 \, \mathrm{mm/s}$ (Fig.\ref{fig:fig7}a1). Then, we may evaluate the time-of-flight of the ballistic sound $t_{sound} = \chi_2/V_{shock} + t_{delay}$ with the delay $t_{delay}= \mathrm{1.2ms}$ and find a sound speed $V_{sound}\sim 80 \, \mathrm{m/s}$. This value of speed is consistent with other acoustic measurements in granular packings under gravity \cite{Liu_1993}, but remains significantly smaller than the front speed at low impact strength (Fig.\ref{fig:fig3}). Fig.\ref{fig:fig7}b shows the waveforms filtered by a high-frequency pass filter ($f_c$ = 8 kHz). While the amplitudes of the relatively low-frequency (or long $\lambda_{shock}$) shock waves reduce as expected, the high-frequency scattered acoustic waves (coda) remain mostly unchanged with a long decaying tail (Figs.\ref{fig:fig7}b1 and b2). The wavelength associated with these coda waves is $\lambda_{sound} = V_s/f_c \sim10 \, \mathrm{mm}\sim3d$. As observed in linear acoustic diffusion experiments \cite{Jia_2004,Brunet_2008}, the addition of the wetting liquid (oil), leads to a very significant decrease in coda waves due to the viscous dissipation that occurs at the grain contacts (Fig.\ref{fig:fig7}b3). 


To make an estimation of viscous and frictional dissipations, we calculate the corresponding intensities of filtered waveforms in Figures \ref{fig:fig7}c. The smoothed intensity profiles (over one realization) are then compared to a diffusion solution in an infinite 3D medium  \cite{Liu_1993} 
\begin{equation}\label{eq:Intensity}
I(t,r) =A t^{-3/2} exp(-B/t) exp(-t/\tau)
\end{equation}
where $A$ is a constant and $B=r^2/4D$ is related to the diffusion coefficient $\mathcal{D}=(1/3)V_sl^*$ with $l^*$ the transport mean free path and $\tau$ is the inelastic absorption time. Taking the distance between the impact and the piezoelectric transducer  $r\sim100 \, \mathrm{mm}$ (considering the initial penetration of the piston) (Fig.~\ref{fig:fig1}), we may infer $\mathcal{D}$ and the quality factor $Q=2\pi f\tau$ by fitting our data. Here we focus on the energy decay rate ($\sim1/Q$), as it is well-defined in the shock regime, with a peak in the coda intensity profile followed by a long detectable tail (Figure \ref{fig:fig7}c). By a best fit of the intensity profiles with Eq. (3), we infer a quality factor $Q\sim2\pi$ in dry packings of glass beads or sand, but a twice smaller$Q\sim\pi$ in wet glass bead packings, confirming an enhanced dissipation due to viscous liquids trapped at particle contacts \cite{Jia_2004,Brunet_2008}.

As to the diffusion coefficient, it is difficult to infer $D$ accurately from the short rising edge of the intensity profiles without enough ensemble average. Nevertheless, we may estimate its order of magnitude $D > 1 \, \mathrm{mm^2/\mu s}$, giving rise to a transport mean free path $l^*>30\, \mathrm{mm}$ $\sim 10{d}$. This value appears much larger than the $\lambda_{sound}\sim3d$ which would be the same order of the scattering mean free path $l_s$ \cite{Jia_2004}. Such a difference might be related to the anisotropic scattering through the chain-like force networks due to our weakly stressed granular packings subject to gravity \cite{Clark_2015}. Further study is needed about this issue.


\section{Conclusion}

We experimentally investigated wave propagation upon impact in 3D granular packings under gravity. Our measurements reveal new features of impulsive shock wave propagation: (i) a transition from linear and weakly nonlinear acoustic waves with sound speed softening at low impact, to strongly nonlinear shock waves at high impact where the shock speed increases with particle-velocity amplitude; and (ii) the characteristic spatial width of shock waves, about ten particle diameters, that remains constant by varying shock amplitudes and is insensitive to the contact microstructure. This is consistent with theoretical 2D frictionless models. 
Finally, our experiments demonstrate the coexistence of two clearly distinct propagation modes: a strongly nonlinear shock wave, and a sound wave composed of a ballistic component followed by a scattered acoustic coda. This sound wave appears after a well-defined time delay relative to the shock front, and its scattered coda (with wavelength ($\lambda\sim 3 d$)) is strongly attenuated in the presence of wetting liquids.

In summary, we propose that shock wave attenuation is governed primarily by geometric spreading and inelastic collisions. Scattering mechanisms play only a minor role, as evidenced by the fact that the saturated shock width ($W\sim 10 d$) is much larger than the particle size $d$. We further suggest that damping follows a cascade process, in which energy is progressively transferred from strongly nonlinear shock waves to weakly nonlinear sound waves through collisions, and ultimately dissipated via viscous absorption of linear and scattered acoustic waves.
The impact propagation problem through such a granular packing offers thus one way of interrogating a 3D FPUT-like system.


\section*{Acknowledgments}
 This work has received support under the program “Investissements d’Avenir” launched by the French
Government. S.W. also acknowledges support by Clervolc Contribution No. XXX. We also thank Martin van Hecke, Surajit Sen for many useful discussions and Rogier van Loo for help with experiments.



\bibliographystyle{unsrt}

\begin{thebibliography}{}
\bibitem{Hecke_2009}
M.\ van Hecke, Jamming of soft particles: geometry, mechanics, scaling and isostaticity, Journal of Physics: Condensed Matter 22 (2009) 033101.

\bibitem{Liu_1998}
A.J.\ Liu, S.R.\ Nagel, Jamming is not just cool any more, Nature 396 (1998) 21-22.

\bibitem{Jaeger_1996}
H.M.\ Jaeger, S.R.\ Nagel, R.P.\ Behringer, Granular solids, liquids, and gases, Reviews of Modern Physics 68 (1996) 1259-1273.

\bibitem{Liu_1992}
C.H.\ Liu, S.R.\ Nagel, Sound in sand,  Phys.\ Rev.\ Let.\ 68 (1992) 2301.

\bibitem{Jia_1999}
X.\ Jia, C.\ Caroli, B.\ Velicky, Ultrasound Propagation in Externally Stressed Granular Media, Phys.\ Rev.\ Let.\ 82 (1999) 1863.

\bibitem{Goddard_1990}
J.D.\ Goddard,Nonlinear elasticity and pressure-dependent wave speeds in granular media,  Proc. R. Soc. London, Ser. A\ 430 (1990) 105.

\bibitem{Johnson_1985}
K.L.\ Johnson, Contact Mechanics, Cambridge University Press, Cambridge (2015).

\bibitem{Nesterenko_1984}
V.F.\ Nesterenko, Propagation of nonlinear compression pulses in granular media, J. Appl. Mech. Tech. Phys. 5 (1984) 733-743.

\bibitem{Nesterenko_2001}
V.F.\ Nesterenko, Dnamics of Heterogeneous Materials, Springer, New-York, 2001.

\bibitem{Coste_1997}
C.\ Coste, E.\ Falcon, S.\ Fauve, Solitary waves in a chain of beads under Hertz contact, Phys.\ Rev. E\ 56 (1997) 6104-9117.

\bibitem{Lazaridi_1985}
A.N.\ Lazaridi, V.F.\ Nesterenko , Observation of a new type of solitary waves in a one-dimensional granular medium, J. of Applied Mechanics and Technical Physics\ 26 (1985) 403-405.

\bibitem{Herbold_2007a}
E.B.\ Herbold, V.F.\ Nesterenko, Shock wave structure in a strongly nonlinear lattice with viscous dissipation, Phys.\ Rev. E\ 75 (2007) 021304.

\bibitem{Herbold_2007b}
E.B.\ Herbold, V.F.\ Nesterenko, Solitary and shock waves in discrete strongly nonlinear double power-law materials, Appl.\ Phys.\ Lett.\ 90 (2007)261902.

\bibitem{Gomez_2012a}
L.R.\ Gómez, A.M.\ Turner, M.\ van Hecke, V.\ Vitelli, Shocks near Jamming, Phys.\ Rev.\ Let.\ 108 (2012) 058001.

\bibitem{Gomez_2012b}
L.R.\ Gómez, A.M.\ Turner, V.\ Vitelli, Uniform shock waves in disordered granular matter, Phys.\ Rev. E\ 86 (2012) 041302.

\bibitem{Wildenberg_2013}
S.\ van den Wildenberg, R.\ van Loo, M.\ van Hecke, Shock Waves in Weakly Compressed Granular Media, Phys.\ Rev.\ Let.\ 111 (2013) 218003.

\bibitem{Clark_2015}
A.H.\ Clark, A.J.\ Petersen, L.\ Kondic, R.P.\ Behringer, Nonlinear force propagation during granular impact, Phys.\ Rev.\ Let.\ 114 (2015) 144502.

\bibitem{Brunet_2008}
T.\ Brunet, X.\ Jia, P.\ Mills, Mechanisms for Acoustic Absorption in Dry and Weakly Wet Granular Media, Phys.\ Rev.\ Let.\ 101 (2008) 138001.

\bibitem{Wildenberg_2015}
S.\ van den Wildenberg, Y.\ Yang, X.\ Jia, Probing the effect of particle shape on the rigidity of jammed granular solids with sound speed measurements, Granular Matter 17 (2015) 419-426.

\bibitem{Manjunath_2012}
M.\ Manjunath, A.P.\ Awasthi, P.H.\ Geubelle, Wave propagation in random granular chains, Phys.\ Rev.\ E\ 85 (2012) 031308.

\bibitem{Jia_2004}
X.\ Jia, Codalike Multiple Scattering of Elastic Waves in Dense Granular Media, Phys.\ Rev.\ Let.\ 93 (2004) 154303.

\bibitem{Liu_1993}
C.\ Liu, S.R.\ Nagel, Sound in a granular material: Disorder and nonlinearity, Phys.\ Rev.\ B \ 48 (1993) 15646-15650.

\bibitem{Jia_2011}
X.\ jia, T.\ Brunet, J.\ Laurent, Elastic weakening of a dense granular pack by acoustic fluidization: Slipping, compaction, and aging, Phys.\ Rev.\ E \ 84 020301(R)(2011)

\bibitem{Sen_2008}
S.\ Sen, J.\ Hong, J.\ Bang, E.\ Avalosa, R.\ Doney, Solitary waves in the granular chain, Physics Reports 462 (2008) 21–66 




















\end{thebibliography}
\end{document}